# The relationship between the number of authors of a publication, its citations and the impact factor of the publishing journal: Evidence from Italy[1]

Giovanni Abramo (corresponding author)
*Laboratory for Studies of Research and Technology Transfer at the*
*Institute for System Analysis and Computer Science (IASI-CNR)*
*National Research Council of Italy*
  ADDRESS: Istituto di Analisi dei Sistemi e Informatica, Consiglio Nazionale delle Ricerche, Via dei Taurini 19, 00185 Roma - ITALY
  tel. +39 06 7716417, fax +39 06 7716461, giovanni.abramo@uniroma2.it

Ciriaco Andrea D'Angelo
*University of Rome "Tor Vergata" - Italy and*
*Laboratory for Studies of Research and Technology Transfer (IASI-CNR).*
  ADDRESS: Dipartimento di Ingegneria dell'Impresa, Università degli Studi di Roma "Tor Vergata", Via del Politecnico 1, 00133 Roma - ITALY
  tel. and fax +39 06 72597362, dangelo@dii.uniroma2.it

**Abstract**

Empirical evidence shows that co-authored publications achieve higher visibility and impact. The aim of the current work is to test for the existence of a similar correlation for Italian publications. We also verify if such correlation differs: i) by subject category and macro-area; ii) by document type; iii) over the course of time. The results confirm world-level evidence, showing a consistent and significant linear growth in the citability of a publication with number of co-authors, in almost all subject categories. The effects are more remarkable in the fields of Social Sciences and Art & Humanities than in the Sciences – a finding not so obvious scrutinizing previous studies. Moreover, our results partly disavow the positive association between number of authors and prestige of the journal, as measured by its impact factor.

---

# 1. Introduction

Collaboration has increased and gained in importance in the domain of scientific research over the last few decades. Various factors are responsible for this, including increasing complexity and interdisciplinarity of science, increasing costs of production factors in research projects, innovations in information and communication technologies. Various studies in the scientometric literature offer empirical evidence that co-authored publications achieve higher visibility and impact. Two world scale studies have been conducted to investigate on the relation between research team size and impact of publications along time. Wuchty et al. (2007) analyzed 19.9 million articles indexed in the Web of Science (WoS) since 1955. Findings show that teams typically produce more frequently cited research than individuals do, and this advantage has been increasing over time. Furthermore, teams now also produce the exceptionally high impact publications, even where that distinction was once the domain of solo authors. Results are detailed for sciences and engineering, social sciences, arts and humanities. Lariviére et al. (2014) extended the period of observation (1900-2011), and the indicators of collaboration (number of authors, number of addresses, and number of countries). The results confirm that an increase in the number of authors leads to an increase in impact, from the beginning of the last century onwards. A similar trend is also observed for the number of addresses and number of countries represented in the byline of an article. The authors note though, that the constant inflation of collaboration since 1900 has resulted in diminishing citation returns: larger teams are necessary to realize higher impact. Recently, observing the 2009-2010 WoS articles and reviews, Waltman and van Eck (2015) confirmed an increasing relation between the number of authors, organizations and countries of a publication and the mean normalized citation score (MNCS) indicator (Waltman et al., 2011). Analyzing the research products submitted by the universities to the first Italian research assessment (VTR, 2006), Franceschet and Costantini (2010) observed a general positive association between cardinality in the byline and the relevant impact (measured by citations) and quality (determined by peer reviewers judgment), notwithstanding notable and interesting counter-examples. Other studies on the subject are in general more limited in scope and make use of sample observations, which makes the generalization of results exposed to the limits and cautions of inferential analysis.

The present work contributes to the existing research, focusing on a single country, Italy, but limiting the observation to the period 2004-2010. Results are compared to those obtained at world level. We investigate also on the relation between team size and impact factor of the journals hosting publications. Furthermore, results are fine-detailed for all 217 subject categories of the WoS core collection. The examination is based on the entire scientific production of all researchers in Italian universities and public research institutions, as indexed in the WoS over 2004-2010. The restriction to Italian data is due to a license agreement between Thomson Reuters and the authors, which limits our access to Italian data only from 2004 onward. We have not investigated more recent periods to achieve robustness of citation counts as a proxy of impact.

In particular, with this work, we intend to test for the Italian case:
- The existence of a correlation between the number of authors of a publication and its impact, measured through the citations received;
- the existence of a correlation between number of authors and the prestige of the journal, measured through its impact factor (IF);

- if such correlations differentiate by subject category (SC) and by macro-area;
- if the relation between the number of authors and the impact of a publication differentiates with the type of document published;
- potential variations over time in the intensity of correlation.

The next section reviews the literature on the subject. Section 3 describes the dataset and the indicators used. Section 4 presents the results of the analyses conducted at the level of subject categories; in Section 5 the analyses are repeated at the aggregate, macro-area level. Sections 6 and 7 explore the relation between number of authors and impact of a publication, first distinguishing by document type and then triennium of observation The paper closes with a summary of the results and the authors' considerations.

## 2. Literature review

The remarkable growth in research collaboration over the past several decades has been the object of numerous studies. Most of these are focused on the analysis of the determinants of scientific collaboration, in a line of research led by Melin & Persson (1996) and Katz & Martin (1997). There are also important studies, but less numerous, attempting to specify a direct functional relationship between the citations received and certain features of the authors list for a scientific article (above all Lariviére et al., 2014; Wuchty et al., 2007; Stewart, 1983).

The choice to collaborate, especially with individuals of different competencies, cultures and experience, is first of all a response to the complexity and interdisciplinarity demanded by certain research themes (He et al., 2009). However, among the determinants of collaboration there can also be considerations that are strictly 'utilitarian': meaning that collaboration is sought out in order to increase the probability of publishing the manuscript (Kalwij and Smit, 2013), of having it accepted by highly ranked journals (Al-Herz et al., 2014), or of receiving citations (Sin, 2011; Leimu et al., 2008). Formal endorsement in the form of co-authorship by a scientist that is already known and well-regarded can gain the manuscript marked advantages in credibility. This phenomenon, known as the 'Matthew effect', was investigated in the pioneering work of Robert K. Merton, in which he indicates the 'effect of cumulativity', meaning that among scientists at parity in quality of publication, the ones that already have more citations will be cited more often (Merton, 1968). Evidently social factors, such as the author's professional standing, play a significant role in citation decisions.

Similar effects could also explain why more prestigious universities have a greater number of collaborations compared to others (Piette and Ross, 1992), and why more advanced nations have a central role in international collaboration networks (Luukkonen et al., 1992). The phenomenon is an important consideration for any principal investigator, called to set up a collaboration team and so the co-authorship of its works. Every co-author has their network of contacts where they are more or less permanently inserted, and which will probably yield citations. From this, more co-authors signifies more social networks, and thus a greater probability of citation for co-authored works.

In addition to the large scale works mentioned in the introduction, various other studies have indeed shown empirically that co-authored publications achieve above-average visibility, measured both in terms of journal importance (Bordons et al., 2013) and citations received (Bordons et al., 2013; Hoekman et al., 2010; Jones et al., 2008), notwithstanding rare exceptions in specific fields of research (Haslam et al., 2008; Didegah and Thelwall, 2013a). Van Dalen & Henken (2001), using a sample of publications appearing in 17 demography journals indexed over 1990-1992 in the Social Science Citation Index, detected that the probability of being cited increased 7% with an increase of one author in the byline. Several years later, using a larger sample, the same authors concluded that the probability of being cited increases 5% with an increase of one author in the byline (Van Dalen & Henken, 2005). Adams et al. (2005), examining data on papers from top U.S. research universities over the period 1981–1999, suggest that output and citations increase with team size (number of co-authors) and that influence (measured by citations) rises with inter-institutional collaboration. In their opinion, increasing team size implies an increase in the division of labor, thus they conclude that productivity increases with the division of scientific labor.

Acedo et al. (2006), in a study of scientific literature on management and organizations, confirm that co-authorship influences the potential impact of an article in the community of reference, meaning in number citations received. This confirms a previous study by Beattie & Goodacre (2004), examining UK and Irish publications in the accounting and finance category, in 1998-1999. An analysis by Skilton (2009), based on a sample of works in top-WoS natural science journals, stresses the fundamental role of 'diversity' in disciplinary backgrounds within the co-author team, and identifies the dominance of 'intellectual' over 'social' capital in citation behavior. Recently, scholars have illustrated how the effect of collaboration on citations tends to diminish if the analysis controls for subtle effects in the composition of the co-author networks and the articles themselves (Hurley et al., 2013; Didegah and Thelwall, 2013b). Finally, we note the correlation between citations received and author numbers could be traced in part to the natural increase in self-citation when works are by more authors (Leimu & Koricheva, 2005) and potentially from distinct institutions (Herbertz, 1995). Although, Lariviére et al. (2014) empirically observe that self-citation contributes to, but does not fully explain, the relationship between impact and collaboration. This relationship seems not a "mechanical" artifact, but rather an effect of the greater epistemic value associated with collaborative research (Beaver, 2004; Wray, 2002).

## 3. Dataset and indicators

The dataset for the analyses consists of the entire 2004-2010 Italian scientific production indexed in the WoS core collection (including Conference Proceedings Citation Indexes and excluding only Chemical and Book indexes).

This is almost 400,000 publications, shown in Table 1 as divided by macro-disciplinary area and document type. For document type, we exclude those not recognizable as true research products (meeting or publication abstracts, editorials, news items, bibliographies; *errata corrige*, etc.)[2].

For evaluation of publication impact we use the citations measured as of 15/05/2014, providing a citation window long enough to ensure robustness of the indicators (Abramo et al., 2011).

---

[2] In column 6, 'other' refers mainly to outputs from the Arts and Humanities, such as exhibitions, production excerpts, film, theatre/film analysis, fiction, prose or poetry, musical scores.

For the evaluation of the prestige of the hosting journal we use IF. Both indicators are expressed in two modes, based on distinct standardization procedures of the absolute value with respect to the relevant distribution:

- $AIR_i$ is the percentile ranking of publication $i$, based on citations of all Italian publications indexed in the same year and SC of publication $i$.
- $JIR_i$ is the percentile ranking of the journal of publication $i$, based on the IF of all journals in the same year and SC of publication $i$. The IF is extracted from the Journal Citation Report® edition of the same year as publication $i$.
- $AII_i$ is the ratio of the citations received by publication $i$, to the average of the distribution of citations received by all cited Italian publications indexed in the same year and SC of publication $i$.[3]
- $JII_i$ is the ratio of the *IF* of the journal of publication $i$, to the average of the distribution of *IF* of all journals in the same year and SC of publication $i$.

For publications in multi-category journals, the value of each indicator is the average value related to each SC.

AIR (JIR) is expressed in percentiles (0 the worst, 100 the best), depending on the cumulative frequency of citations (impact factors) received by publications (journals) of the same year (JCR edition) and subject category. An AIR 20 attributed to publication "i" means that in the same year and subject category we have 80% of publications receiving an equal or higher number of citations than "i". Publications of the same year and subject category and equally cited have the same AIR. An AIR equal to 100 relates to the top cited publication of a given year and subject category. A nihil AIR is attributed to the least cited publication of a given year and subject category. Uncited publications have a nihil AIR.

*Table 1: Italian scientific production per macro-area (2004-2010)*

| Macro-area | No. publications | Articles | Proceedings | Reviews | Letters and other | Number of authors (average) |
|---|---|---|---|---|---|---|
| Art & Humanities | 6,820 | 58% | 12% | 27% | 3% | 1.58 |
| Biology | 63,051 | 86% | 6% | 6% | 1% | 5.88 |
| Biomedical research | 60,266 | 78% | 3% | 12% | 7% | 7.69 |
| Chemistry | 35,608 | 93% | 2% | 4% | 0% | 5.29 |
| Clinical medicine | 99,717 | 80% | 2% | 8% | 10% | 6.99 |
| Earth and space sciences | 25,210 | 85% | 11% | 3% | 0% | 4.83 |
| Economics | 6,982 | 74% | 20% | 6% | 0% | 2.30 |
| Engineering | 91,549 | 59% | 40% | 1% | 0% | 5.31 |
| Law, political and social sciences | 6,641 | 65% | 19% | 12% | 3% | 3.07 |
| Mathematics | 19,960 | 84% | 15% | 1% | 0% | 2.40 |
| Multidisciplinary sciences | 1,609 | 54% | 17% | 23% | 7% | 4.94 |
| Physics | 79,014 | 79% | 19% | 2% | 0% | 14.39 |
| Psychology | 4,290 | 86% | 5% | 7% | 2% | 4.11 |
| Total* | 392,257 | 77% | 13% | 6% | 4% | 7.58 |

\* The total is less than the sum of the column data due to double counts of publications indexed in multi-category journals, falling in different macro-areas.

---

[3] Abramo et al. (2012) demonstrated that the average of the distribution of citations received for all cited publications of the same year and subject category is the most effective scaling factor.

## 4. Analysis at subject category level

The effect of the number of authors on the publication's impact could very easily differentiate by subject category, therefore we begin from this level of analysis, then proceed to the aggregated level of the macro-area. As an example we present the *Neurosciences* subject category, one of the most relevant in terms of size: there are 13,236 Italian publications in the WoS repertories for the period considered. Figure 1 is the box plot of the distribution of number of authors: the median is 6, with average 6.45 and standard deviation 3.92. The maximum value recorded is a publication with 75 authors. This and a number of other outliers can be seen in the points above the upper whisker: all are publications with over 14 authors (409 in all).[4]

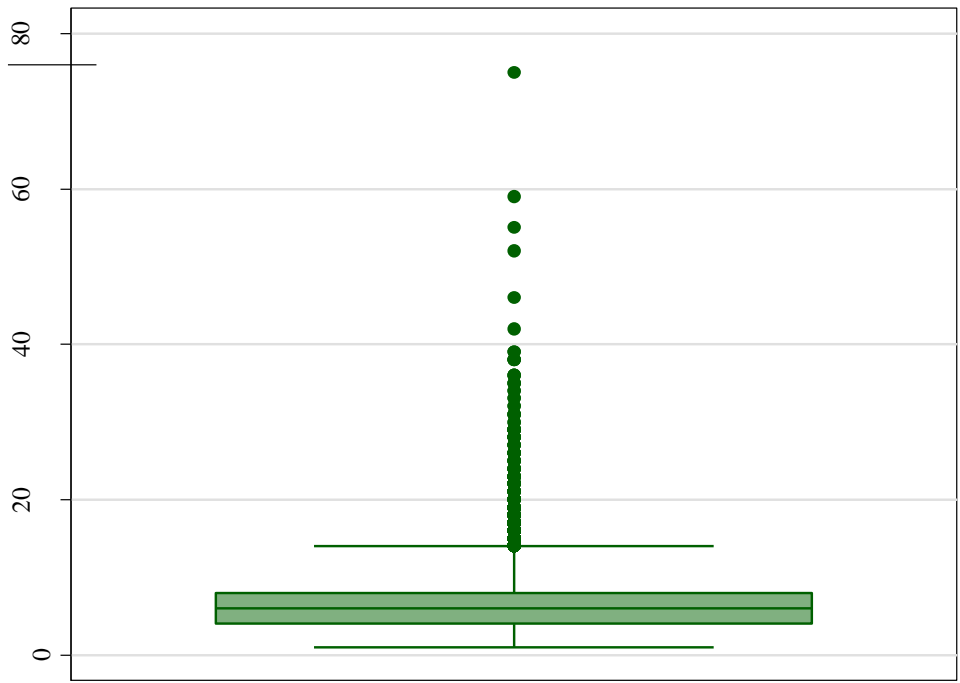

*Figure 1: Box plot of the number of authors of Italian publications in Neurosciences (2004-2010)*

Considering the relation between number of authors and the publication's impact, we reasonably hypothesize that saturation will develop above a certain threshold, beyond which the effect from added co-authors will be marginal. For convention we assume such threshold to be the 95$^{th}$ percentile of the distribution of publications, by number of authors. For *Neurosciences* SC, the convention gives a threshold of exactly 14 authors.[5] In Figure 2 we thus graph the average value of impact of the 13,236 publications in the SC, grouped by number of authors. Measuring impact through AIR and JIR, we note an evident linear trend with R-squared coefficient of determination ($R^2$) around 0.9 for both the indicators. Focusing our attention on the first part of the curve (first four items), there seems to be a logarithmic dependency of impact on the number of authors, with a very marked increment in impact between publications with only one author and those with two authors. In effect, eliminating publications with only one author from the series, and best fitting the $R^2$ to the extrapolated network, the data rise to 0.91 for AIR and 0.95 for JIR, or an almost perfectly linear relation.

---

[4] The upper whisker is equal to Q3+1.5(Q3-Q1).
[5] Above this threshold (distribution right tail) there is a dramatic decrease in observations, and any potential fitting of the data becomes problematic.

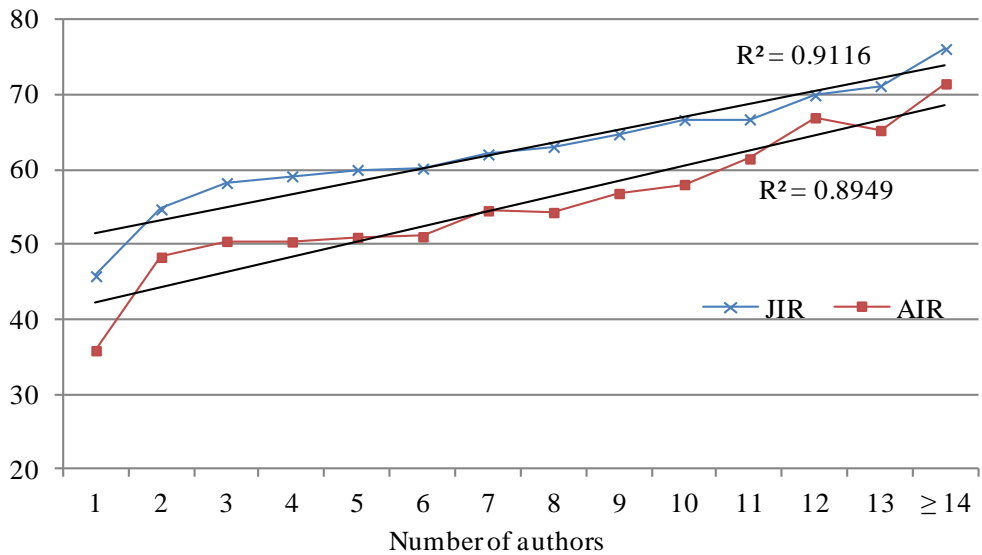

*Figure 2: Average AIR - JIR vs number of authors, for Italian publications in Neurosciences (2004-2010)*

Evaluating impact by AII and JII, we again see a linear dependency between number of authors and impact (Figure 3). However in this case $R^2$ is lower (0.63 for AII and 0.69 for JII). Indeed, the pattern observed for Neurosciences seems quite similar to that observed by Waltman & van Eck (2015) even if their analysis is based on worldwide publications and not restricted to a single subject category. Moreover, extracting the central part of the series (Figure 4), meaning diagramming the average impact only for publications with between two and nine authors, we now observe a convex progression, particularly evident for JII ($R^2 = 0.95$) but also apparent for AII. The greater variability in the trends for these indicators compared to those expressed in percentiles is clearly traceable to the outliers, meaning publications with outstanding values for number of citations and IF. It is well known that distributions of impact indicators are typically highly skewed: the use of the percentile, while compressing the differences between elements with strongly different absolute values of impact, does filter the effects of these outliers. For this, from here forward we conduct the analyses only with AIR and JIR.

In the other SCs the trends are not always so regular as in *Neurosciences*. In *Mathematics, applied* (Figure 5) the number of authors per publication is relatively low (average less than 2.4) and the empirical data concerning the 9,224 publications reveal an anomalous trend: both the number of citations and the IF decrease for publications with more than three authors.

The case of *Optics* is yet more anomalous: the 8,329 publications have a slightly declining impact beginning from triple authorship, while IF is strongly declining beginning from two authors (Figure 6).

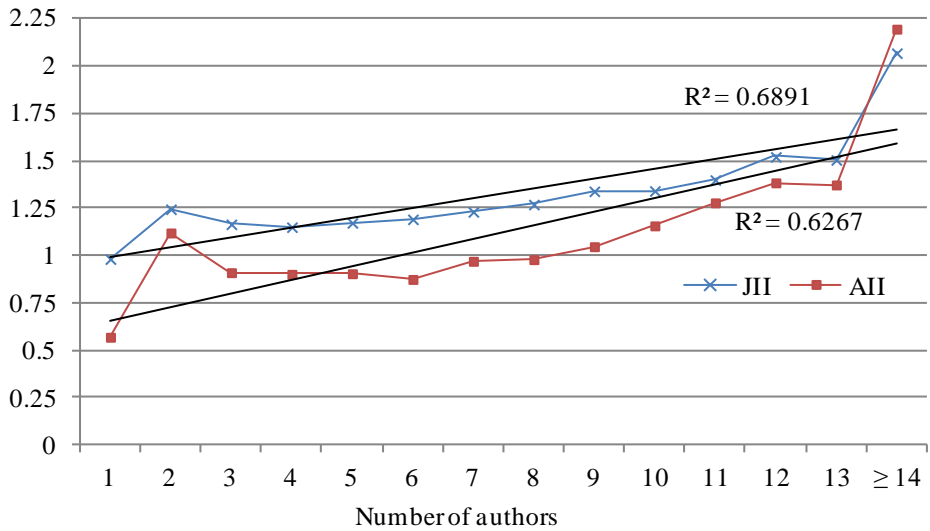

*Figure 3: Average AII - JII vs number of authors, for publications in Neurosciences (2004-2010)*

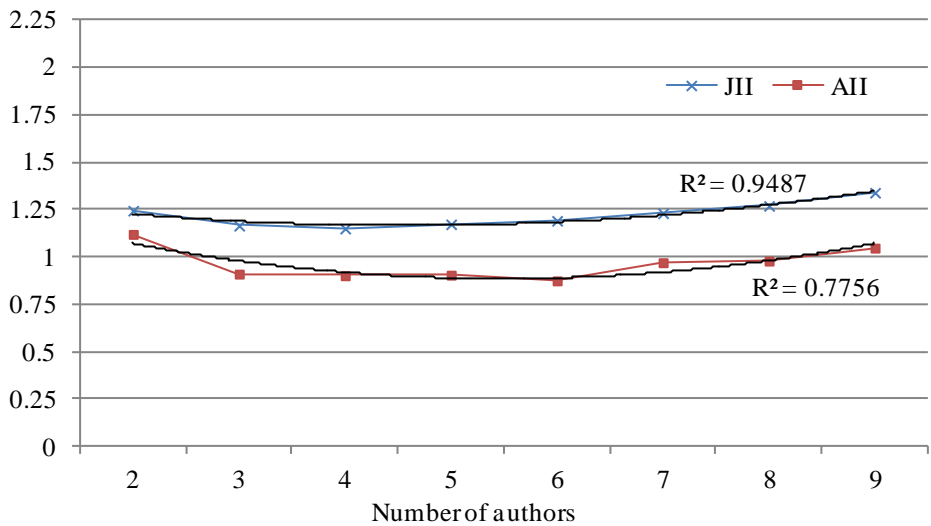

*Figure 4: Average AII - JII vs number of authors (2 to 9), for publications in Neurosciences (2004-2010)*

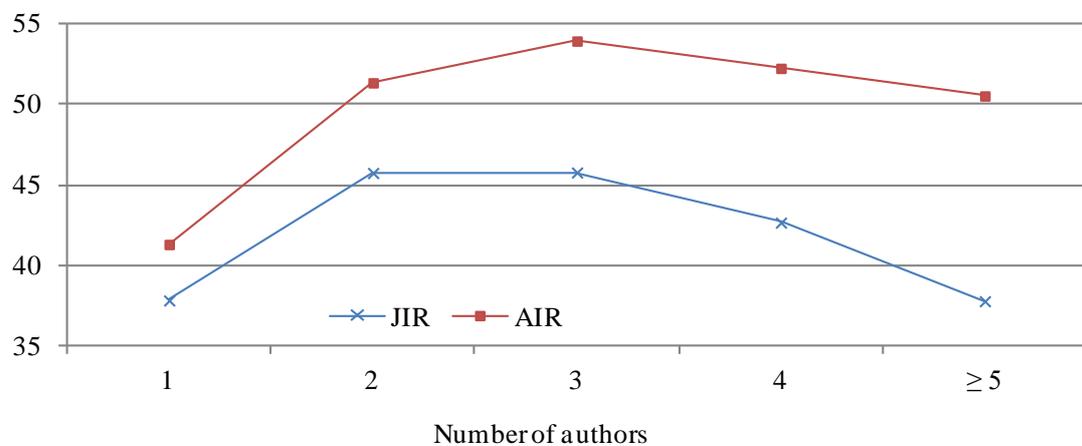

*Figure 5: Average AIR-JIR vs number of authors, for Italian publications in Mathematics-applied (2004-2010)*

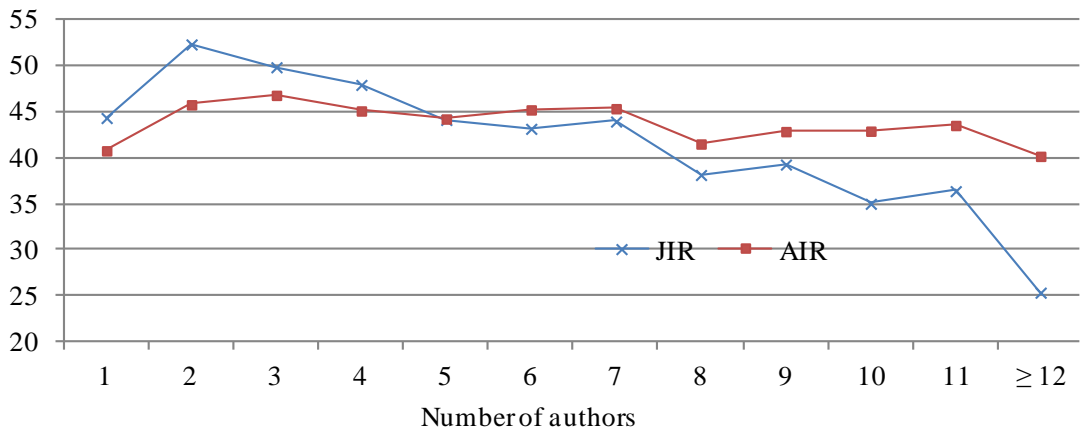

*Figure 6: Average AIR-JIR vs number of authors, for Italian publications in Optics (2004-2010)*

The same analyses were repeated for all 217 WoS SCs[6]: Figure 7 diagrams the dispersion of the coefficients (β) and related $R^2$ for the simple linear regression of AIR versus number of authors.[7] There are only 12 SCs where the regression coefficient is negative, and there is only one case where the fitting is significant ($R^2 \geq 0.5$): these are primarily SCs in Engineering (8) and Physics (3). A different situation emerges from the analyses for JIR: in Figure 8 we note the presence of a full 42 SCs with a negative β value, of which 11 have $R^2$ greater than 0.5. In these categories, the placement for publishing worsens with increasing number of co-authors. In eight of the 11 cases in question, this observation concerns categories of Engineering (6) or Physics (2).

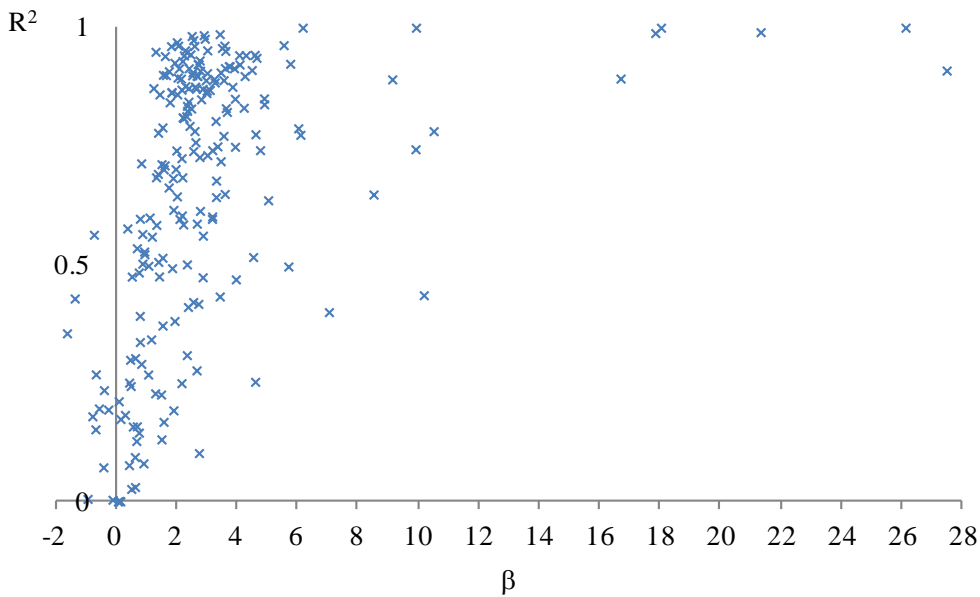

*Figure 7: Distribution of coefficient β and $R^2$, for the linear regression of AIR vs number of authors, in the 217 subject categories analyzed*

---

[6] Indeed, WoS subject category are 251 in all, but for reasons of significance we merged 34 of them with less than 200 Italian publications in the period under observation (mostly SC of Art & Humanities and Social Sciences). The appendix shows the list of SC for each macro areas.

[7] In each regression the independent variable is given by the number of authors, with a limit value that incorporates all the observations equal to or over the 95th percentile. The dependent variable is the average value of the indicator considered.

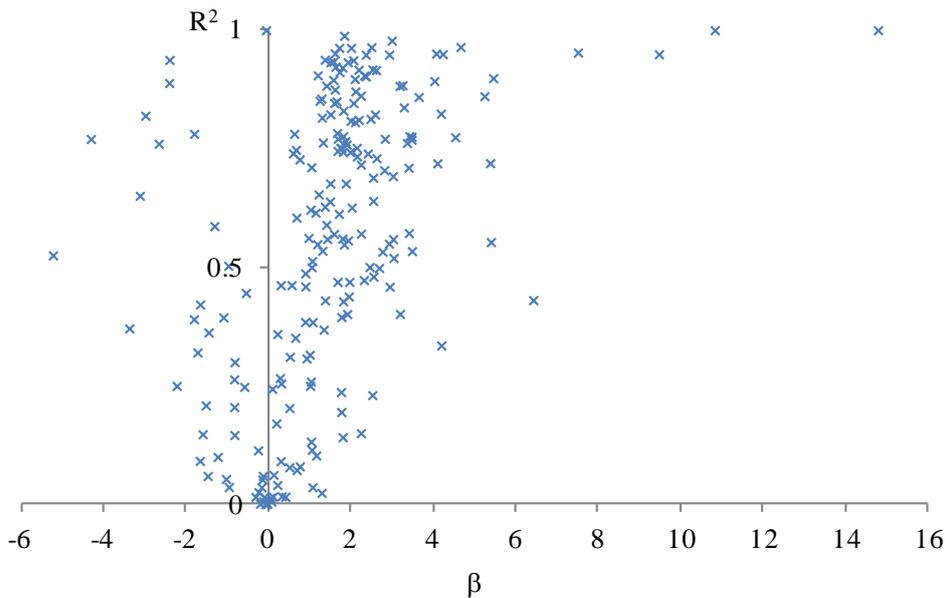

*Figure 8: Distribution of coefficient β and R² for the linear regression of JIR vs number of authors, in the 217 subject categories analyzed*

Table 2 takes the values of the distribution from Figure 7, and for each macro-area provides the descriptive statistics of coefficient β, registered for the subject categories in the linear regression between AIR and number of authors. The macro-areas where the relations between number of authors and AIR is on average strongest are: 1 (Art & Humanities), 9 (Law, political and social sciences) and 13 (Psychology).

*Table 2: Descriptive statistics of β and R² for the linear regression of AIR vs number of authors, in each macro-area*

| Macro-area† | No. subject category | Of which with β<0 | Of which with R²>0.5 | Of which with β>2 | Average β | β variation range |
|---|---|---|---|---|---|---|
| 1 | 12 | 0 | 75.0% | 100.0% | 11.4 | [2.2;27.5] |
| 2 | 29 | 0 | 100.0% | 79.3% | 2.7 | [1.1;4.2] |
| 3 | 14 | 0 | 92.9% | 35.7% | 1.7 | [0.8;2.7] |
| 4 | 8 | 0 | 100.0% | 25.0% | 1.7 | [0.8;2.3] |
| 5 | 40 | 0 | 90.0% | 77.5% | 2.7 | [0.6;4.6] |
| 6 | 12 | 0 | 83.3% | 75.0% | 2.6 | [1.7;3.6] |
| 7 | 8 | 1 | 25.0% | 62.5% | 2.6 | [-0.9;6.0] |
| 8 | 39 | 8 | 43.6% | 30.8% | 1.2 | [-1.6;4.1] |
| 9 | 19 | 0 | 63.2% | 73.7% | 5.1 | [0.5;21.3] |
| 10 | 6 | 0 | 66.7% | 66.7% | 3.0 | [0.1;5.8] |
| 11 | 3 | 0 | 33.3% | 66.7% | 5.2 | [0.8;10.2] |
| 12 | 18 | 3 | 38.9% | 22.2% | 0.9 | [-0.8;3.3] |
| 13 | 9 | 0 | 88.9% | 100.0% | 4.7 | [2.3;17.8] |
| Total | 217 | 12 | 71.9% | 60.8% | 3.0 | [-1.6;27.5] |

† 1=Art & Humanities; 2=Biology; 3=Biomedical research; 4=Chemistry; 5=Clinical medicine; 6=Earth and space sciences; 7=Economics; 8=Engineering; 9=Law, political and social sciences; 10=Mathematics; 11=Multidisciplinary sciences; 12=Physics; 13=Psychology

Table 3 gives a further idea of the strength of this relation, indicating the SC with the highest coefficient β, for each of the macro-areas (having imposed $R^2 \geq 0.5$, and significance of β (p-value of the Fisher's F test less than 0.1) for the linear regression between AIR and number of authors). In Biology, the Mycology SC registers a β value of 4.2, indicating that with increasing number of co-authors (within the interval 1-11) the average marginal increment of AIR is greater than 4%. A similar value is registered in Engineering, environmental, while in Anesthesiology (Medicine macro-area), AIR increases by an average of 4.7% for every additional co-author in the byline. In Mathematics and Economics, this increment is 6%. However the greatest marginal effects are not in the Sciences: in Literature, publications by two or more authors have citability 26% higher than publication by a single author. In History of social sciences an increment in the number of authors (from 1 to 2, and 2 to 3 or more) determines an average increment of 21.3%, while in Psychology, psychoanalysis the increase is 17.8%

*Table 3: Subject categories with the highest β in each macro-area, with regression statistics between AIR and number of authors*

| Macro-area † | Subject category | No. publications | No. authors-average | No. authors-max | No. authors of Top 5% publications per number of authors | β (linear) | $R^2$ | Prob. > F* |
|---|---|---|---|---|---|---|---|---|
| 1 | Literature | 1,274 | 1.1 | 6 | 2 | 26.1 | 1 | 0 |
| 2 | Mycology | 423 | 5.1 | 54 | 11 | 4.2 | 0.831 | 0 |
| 3 | Pathology | 3,712 | 7.3 | 66 | 15 | 2.7 | 0.923 | 0 |
| 4 | Chemistry, multidisciplinary | 5,907 | 5.6 | 34 | 11 | 2.3 | 0.877 | 0 |
| 5 | Anesthesiology | 1,342 | 5.6 | 36 | 11 | 4.6 | 0.943 | 0 |
| 6 | Environmental studies | 888 | 2.6 | 29 | 6 | 3.6 | 0.649 | 0.053 |
| 7 | Economics | 3,972 | 2.1 | 29 | 5 | 6.0 | 0.788 | 0.045 |
| 8 | Engineering, environmental | 2,305 | 4.2 | 43 | 9 | 4.1 | 0.942 | 0 |
| 9 | History of social sciences | 220 | 1.3 | 5 | 3 | 21.3 | 0.991 | 0.062 |
| 10 | Mathematics | 7,672 | 2.1 | 9 | 5 | 5.8 | 0.924 | 0.009 |
| 11 | Education, scientific disciplines | 392 | 2.9 | 22 | 8 | 4.8 | 0.742 | 0.006 |
| 12 | Physics, mathematical | 5,279 | 2.9 | 116 | 7 | 3.3 | 0.803 | 0.006 |
| 13 | Psychology, psychoanalysis | 260 | 1.2 | 5 | 3 | 17.8 | 0.989 | 0.068 |

† 1=Art & Humanities; 2=Biology; 3=Biomedical research; 4=Chemistry; 5=Clinical medicine; 6=Earth and space sciences; 7=Economics; 8=Engineering; 9=Law, political and social sciences; 10=Mathematics; 11=Multidisciplinary sciences; 12=Physics; 13=Psychology
* p value of the Fisher's F test

## 5. Analyses at macro-area and general levels

We now deepen the analysis, aggregating the SCs by macro-area. To do this, we must provide for the fact that the SCs have varying distributions of authors per publication. This occurs even in the same macro-area: Figure 9 shows an example, comparing two SCs of Clinical medicine: Behavioral sciences and Genetics & heredity. The two distributions are clearly different. The values of average (4.6 vs 10.4), mode (3 vs 6), and maximum (25 vs 415) are very distant; Genetics & heredity registers 12% of publications with over 15 co-authors, against 0.4% in Behavioral sciences. Given such phenomena, the independent variable (number of authors of a publication) must be appropriately rescaled, before conducting the aggregate analyses: here we use the decile ranking for number of authors in the relevant SC.

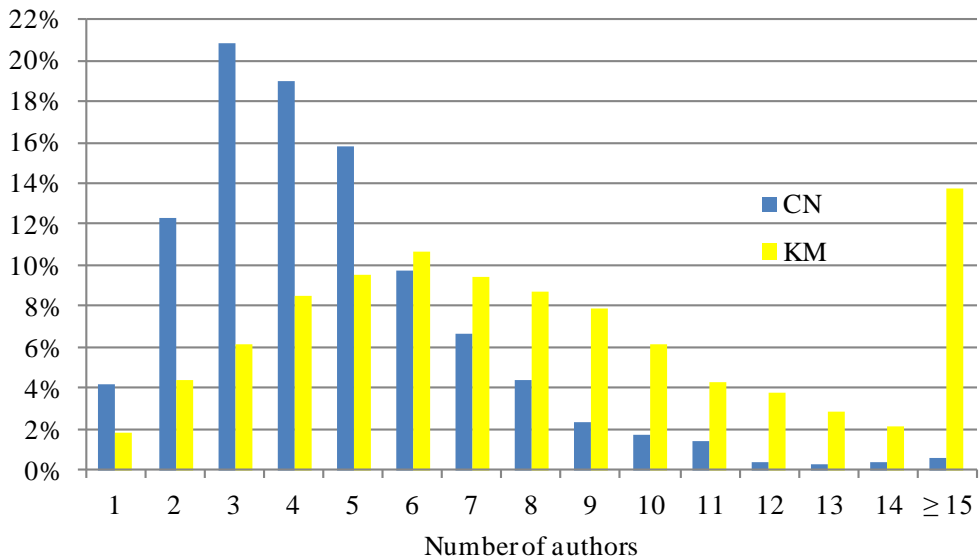

*Figure 9: Frequency distribution of publications per number of authors in Behavioral sciences (CN) and Genetics & heredity (KM)*

Figure 10 presents the results of this operation and the aggregation of data for the SCs in a first cluster of macro-areas (Biology; Biomedical research; Chemistry; Clinical medicine; Earth and space sciences; Mathematics). We observe a very evident trend of increase in the percentile of citations against number of co-authors of publications. As shown in Figure 11, the trend registered for the remaining macro-areas is less evident (Economics; Engineering; Law, political and social sciences; Physics; Psychology). In general we observe increasing impact of the publications with increase in the number of authors but some macro-areas register irregularities, particularly in the first parts of the curves.

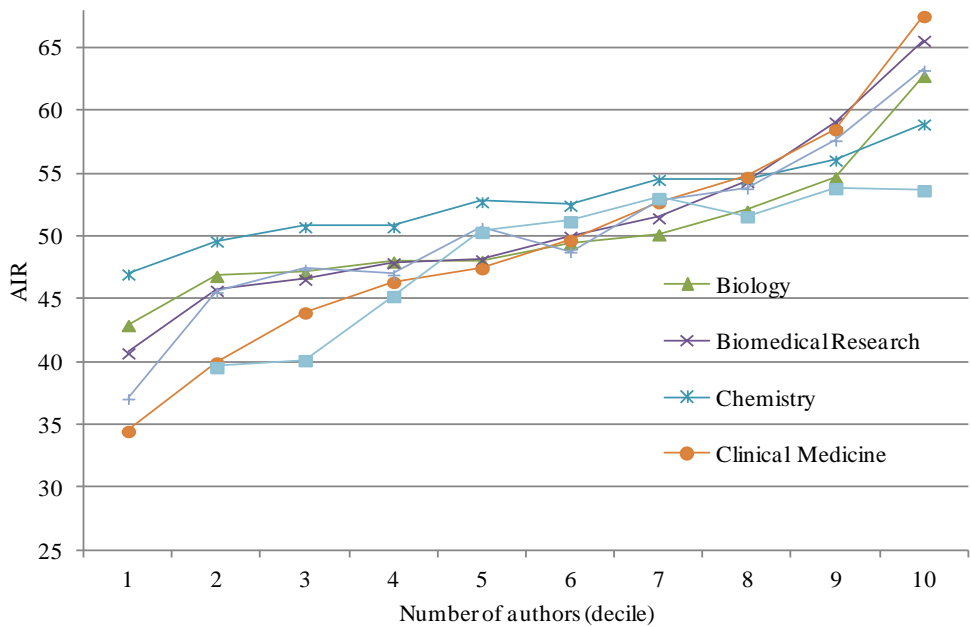

*Figure 10: AIR vs number of authors (in deciles) in six macro-areas*

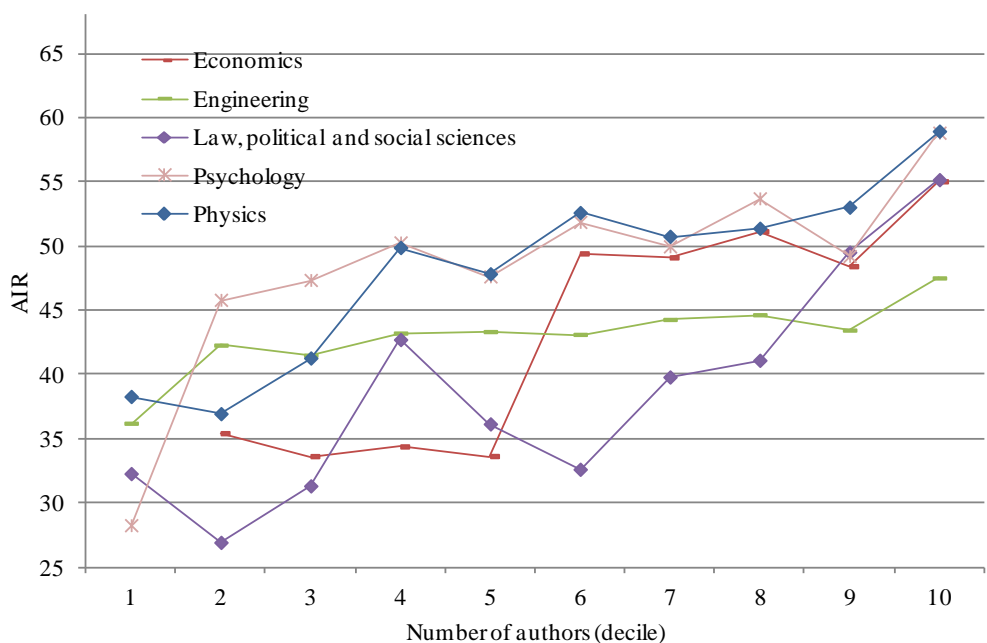

*Figure 11: AIR vs number of authors (in deciles) in five macro-areas*

Table 4 summarizes the results of the regression analyses for the macro-areas, for the impact indicator AIR. In Art & Humanities the number of authors seems not to have any impact on the citability of the publications. Multidisciplinary sciences also seems to be a unique case: while the regression coefficient is still very high, it results as not significant.[8] In Engineering the influence of number of authors on impact is modest (the only macro-area with a value of regression coefficient below one). In Chemistry as well the impact is little influenced by number of authors ($\beta = 1.106$). In contrast, Clinical medicine registers the highest $\beta$ (3.044). Below this 'top' macro-area the next ones are Economics (2.866) and Law, political and social sciences (2.431). This confirms the previous observation, that the link between citability and number of co-authors is more evident in the Social Sciences than the Sciences. Even in a general analysis without any consideration of discipline (last line of Table 4) the relation between number of authors and impact of a publication is significant and monotonically increasing ($\beta = 1.692$ and $R^2 = 0.959$).

Table 5 presents the results for the same regression analysis with JIR as the dependent variable.[9] Other than the anomaly of Multidisciplinary sciences, also seen earlier, we note that in Law, political and social sciences, Engineering and Mathematics, the regressions do not yield significant results ($R^2$ low and p-value of the Fisher's F test greater than 0.1). On contrast, in the life sciences (Biology; Biomedical research; Clinical medicine), increasing number of co-authors is accompanied by increasing prestige of the publishing journal. In Economics, the average gain in IF from one decile to the next in the distribution of number of authors is 2.2%; in Psychology the gain per decile is 1.9%, and in Physics 1.7%.

---

[8] For this, in Figures 10 and 11 we omit the series for these two macro-areas.
[9] The table does not include Art & Humanities, since there is no Journal Citation Report for the macro-area (journals are not assigned an impact factor). From here on, the macro-area is excluded from analysis and discussion.

*Table 4: Linear regression of AIR vs number of authors (in deciles) per macro-area*

| Macro-area | β | $R^2$ | Prob > F* |
|---|---|---|---|
| Art & Humanities | -0.038 | 0.000 | 0.982 |
| Biology | 1.613 | 0.808 | 0.000 |
| Biomedical research | 2.236 | 0.893 | 0 |
| Chemistry | 1.106 | 0.948 | 0 |
| Clinical medicine | 3.044 | 0.950 | 0 |
| Earth and space sciences | 2.229 | 0.893 | 0 |
| Economics | 2.866 | 0.787 | 0.001 |
| Engineering | 0.781 | 0.678 | 0.003 |
| Law, political and social sciences | 2.431 | 0.706 | 0.002 |
| Mathematics | 1.891 | 0.821 | 0.002 |
| Multidisciplinary sciences | 2.997 | 0.231 | 0.190 |
| Physics | 2.162 | 0.854 | 0.000 |
| Psychology | 2.026 | 0.594 | 0.009 |
| Total | 1.692 | 0.959 | 0 |

*\* p value of the Fisher's F test*

*Table 5: Linear regression of JIR vs number of authors (in deciles) per macro-area*

| Macro-area | β | $R^2$ | Prob > F* |
|---|---|---|---|
| Biology | 1.760 | 0.899 | 0 |
| Biomedical research | 1.601 | 0.828 | 0.000 |
| Chemistry | 0.895 | 0.616 | 0.007 |
| Clinical medicine | 1.663 | 0.846 | 0.000 |
| Earth and space sciences | 1.596 | 0.837 | 0.000 |
| Economics | 2.201 | 0.567 | 0.019 |
| Engineering | -0.129 | 0.005 | 0.844 |
| Law, political and social sciences | -0.248 | 0.016 | 0.727 |
| Mathematics | 0.717 | 0.360 | 0.116 |
| Multidisciplinary sciences | -4.809 | 0.383 | 0.076 |
| Physics | 1.721 | 0.526 | 0.018 |
| Psychology | 1.881 | 0.717 | 0.002 |
| Total | 0.712 | 0.598 | 0.009 |

*\* p value of the Fisher's F test*

The Physics macro-area is in fact known for its extensive international collaborations, particularly in high-energy physics, with research results codified by hundreds and even thousands of authors in articles that typically appear in top journals. However the relation between number of authors and JIR seems less evident in zones far from the right tail of distribution of the independent variable, which could explain the $R^2$ of 0.526: still certainly significant but not as high as in other macro-areas. The last line of Table 5 attests that the relation between number of authors and prestige of the publishing journal is one of constant increase, but with a first derivative (and level of significance) clearly less than observed for the citations.

### 6. Influence of the document type

Review articles aim at reviewing the scientific literature on a particular topic, while research articles present new results and conference papers in general intermediate results of in progress research. Such peculiarities should be reflected in observable features of publications. To this purpose Barrios et al. (2013) investigated similarities and differences between different document types in Psychology, in terms of impact and some structural features including number of authors and affiliations. Before that, Sin (2011) had carried out a similar analysis in the field of Library and information science.

We may expect then that the relationship between the number of authors of a publication and its impact is somehow influenced by the document type of the publication itself.

Sorting the publications by document type and repeating the regressions proposed in the previous section, we have the results seen in Table 6. In the regression for AIR vs number of authors,[10] the coefficients remain positive and significant for 'articles', in all the macro-areas. The situation changes for 'conference proceedings': the coefficients of regression lose significance in seven macro-areas (Biology; Chemistry; Clinical medicine; Economics; Law, political and social sciences; Mathematics; Psychology); the coefficients are positive and significant in four other areas (Earth and space sciences; Engineering; Multidisciplinary sciences; Physics), but their values are halved compared to 'articles'. In Biomedical research the coefficient is actually negative (-0.686), indicating that in this macro-area, conference presentations by many authors are penalized in terms of citability, compared to those by few authors.

The analysis for 'reviews' yields an interesting observation: in a full five macro-areas (Earth and space sciences; Economics; Engineering; Mathematics; Physics), not only is the relation between citations and number of authors significant, but it is stronger than for articles. In Economics, the regression coefficient β for reviews reaches 11.52, compared to 1.544 for articles; in Mathematics, β is 6.522 for reviews and 1.716 for articles. We recall that these two macro-areas are also among those with the lowest average authors per publication, from the last column of Table 1. It should be noted that in the Social Sciences, results could be partly distorted because of WoS misclassification of journal articles containing original research into the "review" or "proceedings paper" category (Harzing, 2013).

*Table 6: Linear regression of AIR vs number of authors (in deciles) per macro-area and document type*

| Macro-area† | Articles | | Proceedings | | Reviews | | Letters and other | |
|---|---|---|---|---|---|---|---|---|
| | β | $R^2$ | β | $R^2$ | β | $R^2$ | β | $R^2$ |
| 2 | 1.903*** | 0.872 | 0.348 | 0.224 | 0.672* | 0.354 | 3.053*** | 0.856 |
| 3 | 2.317*** | 0.904 | -0.686*** | 0.701 | 0.790* | 0.323 | 2.666*** | 0.885 |
| 4 | 1.405*** | 0.908 | 0.318 | 0.139 | -0.291 | 0.195 | 3.247* | 0.435 |
| 5 | 2.505*** | 0.931 | 0.071 | 0.020 | 1.189*** | 0.630 | 2.035*** | 0.892 |
| 6 | 1.740*** | 0.849 | 0.950** | 0.539 | 2.729** | 0.584 | 2.177*** | 0.668 |
| 7 | 1.544* | 0.343 | 0.426 | 0.162 | 11.52*** | 0.766 | 8.230* | 0.734 |
| 8 | 1.287*** | 0.745 | 0.772** | 0.446 | 1.376*** | 0.684 | 2.242** | 0.559 |
| 9 | 2.203*** | 0.849 | -0.328 | 0.049 | 2.786 | 0.103 | 2.076** | 0.477 |
| 10 | 1.716*** | 0.702 | 1.637 | 0.048 | 6.522*** | 0.853 | 4.336*** | 0.874 |
| 11 | 4.032** | 0.592 | 2.044** | 0.843 | 2.344 | 0.057 | 2.056 | 0.226 |
| 12 | 1.793*** | 0.832 | 0.639*** | 0.743 | 2.300*** | 0.727 | 3.835*** | 0.731 |
| 13 | 1.817*** | 0.748 | 1.113 | 0.248 | 1.541 | 0.046 | -0.733 | 0.036 |
| Total | 2.091*** | 0.950 | 0.858*** | 0.600 | 0.470 | 0.045 | 2.579*** | 0.933 |

*Statistical significance: \*p-value <0.10, \*\*p-value <0.05, \*\*\*p-value <0.01*

*† 2=Biology; 3=Biomedical research; 4=Chemistry; 5=Clinical medicine; 6=Earth and space sciences; 7=Economics; 8=Engineering; 9=Law, political and social sciences; 10=Mathematics; 11=Multidisciplinary sciences; 12=Physics; 13=Psychology*

---

[10] For the type of analyses in play, that for JIR would be of little use, particularly since proceedings papers are not assigned an impact factor.

# 7. An inter-temporal analysis

We now test whether the relation between number of authors and impact of a publication varies over time, at least over the brief period of the observed publication window (2004-2010). Given the trend of significant increase in the practice of collaboration over time (Abramo et al., 2013; Uddin et al., 2012; Schmoch and Schubert, 2008; Abt, 2007; Glänzel et al., 1999), we would expect just such an increase in the length of byline for Italian publications over the period. To verify, we subdivide the publication window in two distinct triennia: 2004-2006 and 2008-2010, and calculate, for each SC, the average number of authors per publication in the two triennia. The results of the analysis are presented in Table 7: in Chemistry, Earth and space sciences, Economics and Mathematics, the average number of authors per publication increases in every subject category. In all the other macro-areas but one, the average number of authors per publication increases in not less than two-thirds of the subject categories. The sole exception is the macro-area of Psychology: in five of the nine subject categories, the average length of the byline decreases. The extreme case is the Psychology, social SC: the 92 publication over the 2004-2006 triennium have an average of 7.5 authors, against 3.97 (-47.3%) for 174 publications in the 2008-2010 period. In reality the observation is highly conditioned by the outliers: in the first period there were four publications with 69 to 131 co-authors, while in the second period the 'top collaboration' had a byline of only 61 authors.

The situation is different for the Engineering, aerospace SC: publications in the first triennium show an average number of co-authors of 6.06, declining to 4.2 in the second triennium (-30.7%), a result that does not vary with exclusion of the outliers. However these specific SCs are exceptions: in 175 of the 205 categories investigated (85.4%), the average number of authors per publication increases between successive triennia: in Genetics & heredity the difference is greater than 50%; in Instruments & instrumentation (a large Engineering SC with over 1,000 publications per year), the average number of authors per publication increases from 8.7 in 2004-2006 to 31.7 (+263.8%) in 2008-2010 − attributed above all to the tripling of publications with over 100 authors.

*Table 7: Descriptive statistics of the variation in average number of authors in each macro-area, between 2004-2006 and 2008-2010*

| Macro-area | No. subject categories | Of which with increasing average co-authors per paper | Average increase (%) | Variation range of average co-authors per paper (%) |
|---|---|---|---|---|
| Biology | 29 | 27 (93.1%) | 11.7 | [-0.6; 28.4] |
| Biomedical research | 14 | 12 (85.7%) | 7.8 | [-3.4; 21.9] |
| Chemistry | 8 | 8 (100%) | 7.7 | [4.4; 11.4] |
| Clinical medicine | 40 | 32 (80%) | 7.9 | [-14.3; 50.2] |
| Earth and space sciences | 12 | 12 (100%) | 11.5 | [4.3; 23.4] |
| Economics | 8 | 8 (100%) | 12.8 | [7.1; 19.4] |
| Engineering | 39 | 31 (79.5%) | 9.6 | [-30.7; 263.8] |
| Law, political and social sciences | 19 | 16 (84.2%) | 16.9 | [-5.3; 68.2] |
| Mathematics | 6 | 6 (100%) | 7.0 | [3.5; 20.4] |
| Multidisciplinary sciences | 3 | 2 (66.7%) | 20.2 | [-4.7; 49.8] |
| Physics | 18 | 17 (94.4%) | 8.4 | [-1.1; 33.8] |
| Psychology | 9 | 4 (44.4%) | -1.7 | [-47.3; 26.0] |
| Total | 205 | 175 (85.4%) | 10.0 | [-47.3; 263.8] |

We can thus confirm that the average number of authors per publication has grown significantly between the two triennia, in the overwhelming majority of SCs. Next we verify if there is a change in the dependence of relative impact of a publication on the length of its byline. Figure 12 presents the trend of AIR in function of number of authors (expressed as deciles in the distribution of SC) for all Italian publications, distinguished by triennia. The two curves show a linear fitting that is practically identical, with β coefficients of 1.96 for 2004-2006, and 1.99 for 2008-2010, and $R^2$ respectively equal to 0.944 and 0.928. The fact that the curve for the publications of the 2008-2010 triennium is almost always under that of the first biennium should not distract us: AIR, although based on a rescaling of citations accounting for year of publication, is still sensitive to the incidence of non-cited publications, which are less numerous for the first triennium due to the longer citation window (citations for both triennia are counted as of 15/05/2014).

The identity of linear fitting does not occur for JIR, since in this case the impact indicator is linked to the *IF* of the journal as registered year for year: Figure 13 shows the analyses based on this second indicator. It shows a highly irregular and oscillating trend, generally similar in the two triennia considered.

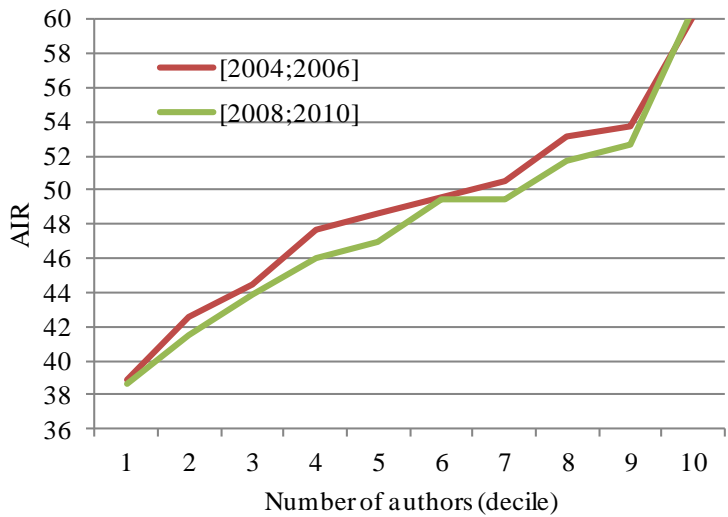

*Figure 12: AIR vs number of authors (in deciles) for all Italian publications*

In reality, the analyses conducted at the level of the single macro-area reveal a certain differentiation over time in the link between impact and number of authors (Table 8). In Engineering this link is practically insignificant in the 2004-2006 triennium, while becoming significant in the next triennium. However in the life sciences, the link seems to weaken: in Biology, Biomedical research and Clinical medicine, the linear regression coefficient β diminishes in the second triennium although not by much. The same occurs in more pronounced manner in Economics (β from 3.91 to 2.268) and in Law, political and social sciences (β from 2.742 to 2.297).

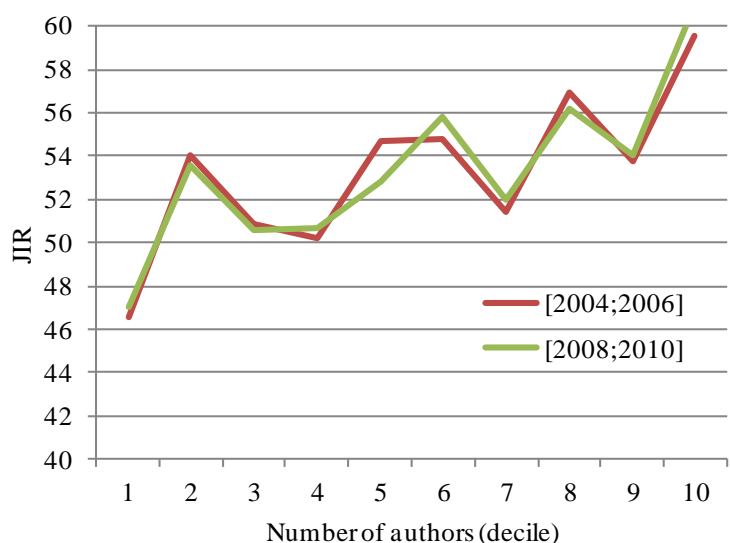

*Figure 13: JIR vs number of authors (in deciles) for all Italian publications*

*Table 8: Linear regression of AIR vs number of authors (in deciles) per macro-area and triennium*

| Macro-area | [2004-2006] | | | [2008-2010] | | |
|---|---|---|---|---|---|---|
| | β | $R^2$ | Prob > F* | β | $R^2$ | Prob > F* |
| Biology | 1.729 | 0.840 | 0.000 | 1.546 | 0.763 | 0.001 |
| Biomedical research | 2.417 | 0.933 | 0 | 2.104 | 0.847 | 0.000 |
| Chemistry | 1.070 | 0.851 | 0.000 | 1.151 | 0.945 | 0 |
| Clinical medicine | 3.191 | 0.967 | 0 | 2.881 | 0.921 | 0 |
| Earth and space sciences | 1.960 | 0.847 | 0.000 | 2.473 | 0.918 | 0 |
| Economics | 3.910 | 0.676 | 0.007 | 2.268 | 0.743 | 0.003 |
| Engineering | 0.502 | 0.300 | 0.101 | 0.991 | 0.758 | 0.001 |
| Law, political and social sciences | 2.742 | 0.771 | 0.001 | 2.297 | 0.568 | 0.012 |
| Mathematics | 1.916 | 0.547 | 0.036 | 1.894 | 0.804 | 0.003 |
| Multidisciplinary sciences | 2.506 | 0.171 | 0.269 | 3.243 | 0.269 | 0.152 |
| Physics | 1.853 | 0.799 | 0.001 | 2.453 | 0.891 | 0 |
| Psychology | 2.468 | 0.793 | 0.001 | 1.959 | 0.480 | 0.026 |

*\* p value of the Fisher's F test*

In contrast, in Earth and space sciences and in Physics the link seems reinforced, witnessed by an increase in the linear regression coefficients of 26% and 32%, respectively. Figures 14 and 15 show the data plot for these two macro-areas, bringing out that the right section of the curve shows the variation with greatest slope. In Physics in particular, we observe a clear separation between the two curves beginning from the seventh decile. Once again, the publications of the second triennia are evaluated over a shorter citation window, and we thus expect a greater incidence of non-cited publications. Instead, the data reveal the exact opposite. The incidence of publications with many authors (seventh decile and up) that are left un-cited is greater for the first triennium than the second. We might suspect that the phenomenon is due to the higher deciles featuring ever longer bylines in the second triennium, compared to the first, but the data again indicate the exact opposite. In Physics in the 2004-2006 triennium the publications of the eighth decile have an average number of co-authors of 6.1; the ninth decile has an average of 11.0, and the tenth decile has 100.5. In the second triennium these values descend respectively to 5.9, 10.3 and 96.8. In Earth and space sciences we see the same effect, at least for the top two deciles, which are the ones where see a significant shift in AIR. The increase in slope between the final part of the curve could thus result from an increase over time in the so-called 'immediacy' of the works by many authors, meaning the average value of the citations received immediately following the appearance of a publication (Marton, 1985). In other words, in Physics (less so in Earth and space sciences), the number of co-authors has a significant effect both on absolute impact of a publication and on its *immediacy*, and this effect is increasing over time.

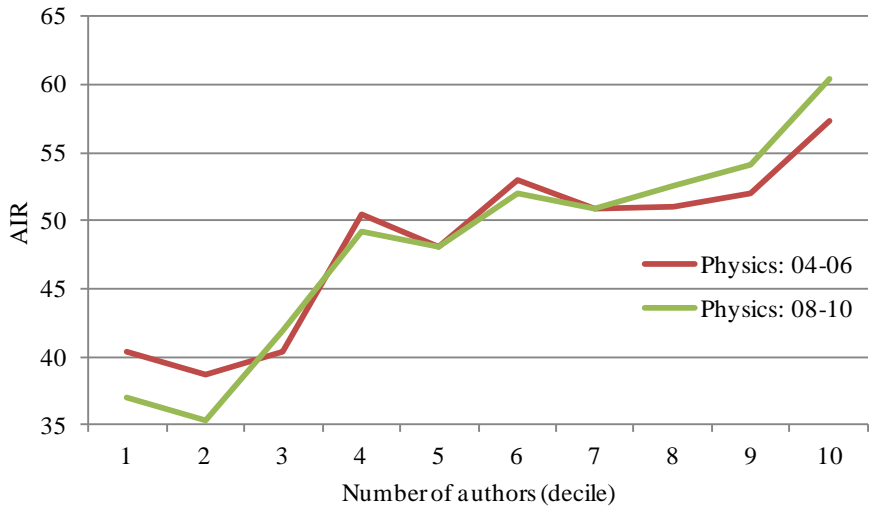
*Figure 14: AIR vs number of authors (in deciles) for publications in Physics*

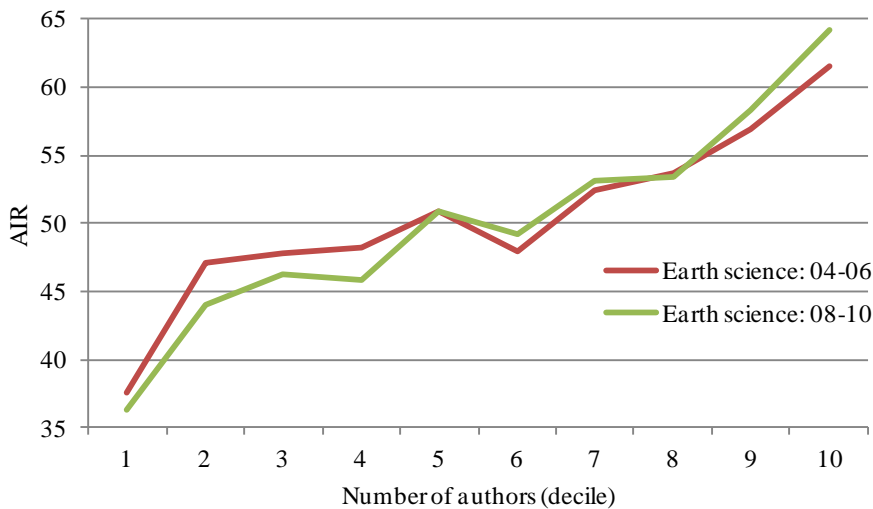
*Figure 15: AIR vs number of authors (in deciles) for publications in Earth and space sciences*

## 8. Conclusions

The literature is generally unanimous in recognizing that co-authored publications achieve above-average visibility and impact. This country-level work has attempted to confront findings for Italy with world-level ones. In particular we investigated on the correlation between the number of the publication co-authors, citations received, and IF of the publishing journal, for publications with at least one Italian institution, in all of the 217 WoS subject categories.

The results confirm world-level evidence, showing a consistent and significant linear growth in the citability of a publication with number of co-authors, in almost all subject categories. The effects are more remarkable in the fields of Social Sciences and Art & Humanities than in the Sciences – a finding not so obvious scrutinizing previous studies. It must be noted though, that the WoS coverage of overall publications in both the Social Sciences and the Art & Humanities is relatively low, therefore the significance of our results in those disciplines should be interpreted with caution.

Moreover, our results partly disavow the positive association between number of authors and prestige of the journal, as measured by its impact factor: in many subject categories the relation between number of co-authors and prestige of the hosting journal is not significant, and in some cases is negative. The stratification by "document type" offers additional insights: for conference proceedings, the correlation between number of authors and citations is weaker than for articles. As a matter of fact in most macro-areas, the correlation for conference proceedings (co-authors vs citations) is not significant, and in one case (Biomedical research) it is even negative. In contrast, in Earth and space sciences, Economics, Engineering, Mathematics and Physics, the regressions for 'reviews' show citations significantly increasing with number of authors: an increase that is even greater than for articles.

The inter-temporal analyses confirm prior literature: although focusing on a short publication period, the average number of co-authors for publications has increased. This occurred in 85% of the subject categories investigated. Differently from what observed by Wuchty et al. (2007), over time the link between number of authors and citation counts of the publications seems affected by differing trends in the various macro-areas: in the Life sciences the link weakens over time, while it strengthens in Physics and Earth and space sciences, perhaps because of increasing 'immediacy' over the two periods. In general, the empirical results are still open to interpretation. It is undeniable that the increasing complexity and interdisciplinarity of research makes it necessary to resort to collaboration, and draw on the various competencies available in a research team to confront the ever-more demanding challenges. It is thus natural that the quality of a scientific work would be linked to the qualitative-quantitative composition of the research team that produces it (Beaver, 2004; Wray, 2002). Still, the knowledge of a 'signaling' effect, in which long bylines gain visibility for publications, also creates strictly opportunistic incentives for collaboration. Whatever is the fundamental determinant, the analyses show that the correlation to increasing authors is very strong for citations, but less so in terms of the impact for the publishing journal (IF).

This last observation stimulates a consideration for further research: if we clustered all the journals on the basis of IF, would we see significant variations in the average number of co-authors per publication, related to these 'IF clusters'? And would the correlation between the number of co-authors and IF vary across clusters?

Finally, one could also elaborate the regression analyses by including independent variables concerning the scientific profile of the co-authors: could we then verify the existence of the Matthew effect, and if it exists, how much does the effect reduce the incidence of total number of co-authors on the publication's citability?

**Appendix – list of subject categories (SC)**

| Macro-area | Subject categories |
|---|---|
| Art & Humanities | Archaeology; Architecture & Art; Art; Classics; Dance, Theater, Music, Film and Folklore; History; Humanities, Multidisciplinary; Language & Linguistics; Literature; Medieval & Renaissance Studies; Philosophy; Religion |
| Biology | Agricultural Engineering; Agriculture, Dairy & Animal Science; Agriculture, Multidisciplinary; Agronomy; Biochemical Research Methods; Biochemistry & Molecular Biology; Biodiversity Conservation; Biology; Biophysics; Biotechnology & Applied Microbiology; Cell & Tissue Engineering; Cell Biology; Developmental Biology; Ecology; Entomology; Evolutionary Biology; Fisheries; Food Science & Technology; Forestry; Horticulture; Marine & Freshwater Biology; Mathematical & Computational Biology; Microbiology; Mycology; Plant Sciences; Reproductive Biology; Soil Science; Veterinary Sciences; Zoology |
| Biomedical Research | Allergy; Anatomy & Morphology; Chemistry, Medicinal; Hematology; Immunology; Infectious Diseases; Medical Laboratory Technology; Medicine, Research & Experimental; Oncology; Pathology; Pharmacology & Pharmacy; Radiology, Nuclear Medicine & Medical Imaging; Toxicology; Virology |
| Chemistry | Chemistry, Analytical; Chemistry, Applied; Chemistry, Inorganic & Nuclear; Chemistry, Multidisciplinary; Chemistry, Organic; Chemistry, Physical; Electrochemistry; Polymer Science |
| Clinical Medicine | Andrology; Anesthesiology; Audiology & Speech-Language Pathology; Behavioral Sciences; Cardiac & Cardiovascular Systems; Clinical Neurology; Critical Care Medicine; Dentistry, Oral Surgery & Medicine; Dermatology; Emergency Medicine; Endocrinology & Metabolism; Gastroenterology & Hepatology; Genetics & Heredity; Geriatrics & Gerontology; Health Care Sciences & Services; Integrative & Complementary Medicine; Medicine, General & Internal; Medicine, Legal; Neuroimaging; Neurosciences; Nutrition & Dietetics; Obstetrics & Gynecology; Ophthalmology; Orthopedics; Otorhinolaryngology; Parasitology; Pediatrics; Peripheral Vascular Disease; Physiology; Psychiatry; Public, Environmental & Occupational Health; Rehabilitation; Respiratory System; Rheumatology; Sport Sciences; Substance Abuse; Surgery; Transplantation; Tropical Medicine; Urology & Nephrology |
| Earth & Space sciences | Environmental Sciences; Environmental Studies; Geochemistry & Geophysics; Geography, Physical; Geology; Geosciences, Multidisciplinary; Limnology; Meteorology & Atmospheric Sciences; Mineralogy; Oceanography; Paleontology; Water Resources |
| Economics | Business; Business, Finance; Economics; Information Science & Library Science; Management; Planning & Development; Public Administration; Transportation |
| Engineering | Automation & Control Systems; Computer Science, Artificial Intelligence; Computer Science, Cybernetics; Computer Science, Hardware & Architecture; Computer Science, Information Systems; Computer Science, Interdisciplinary Applications; Computer Science, Software Engineering; Computer Science, Theory & Methods; Construction & Building Technology; Engineering, Aerospace; Engineering, Biomedical; Engineering, Chemical; Engineering, Civil; Engineering, Electrical & Electronic; Engineering, Environmental; Engineering, Geological; Engineering, Industrial; Engineering, Manufacturing; Engineering, Marine; Engineering, Mechanical; Engineering, Multidisciplinary; Engineering, Ocean and Marine; Instruments & Instrumentation; Materials Science, Biomaterials; Materials Science, Ceramics; Materials Science, Characterization & Testing; Materials Science, Coatings & Films; Materials Science, Composites; Materials Science, Multidisciplinary; Materials Science, Textiles, Paper & Wood; Medical Informatics; Metallurgy & Metallurgical Engineering; Mining & Mineral Processing; Nanoscience & Nanotechnology; Nuclear Science & Technology; Remote Sensing; Robotics; Telecommunications; Transportation Science & Technology |
| Law, political & social sciences | Anthropology; Area Studies; Communication; Education & Educational Research; Ethics; Geography; Gerontology; Health Policy & Services; History of Social Sciences; International Relations; Law; Nursing; Political Science; Social Issues, Multidisciplinary; Social Sciences, Biomedical; Social Sciences, Interdisciplinary; Social Sciences, Mathematical Methods; Sociology; Urban Studies |
| Mathematics | Logic; Mathematics; Mathematics, Applied; Mathematics, Interdisciplinary Applications; Operations Research & Management Science; Statistics & Probability |
| Multidiscipl. sciences | Education, Scientific Disciplines; History & Philosophy of Science; Multidisciplinary Sciences |
| Physics | Acoustics; Astronomy & Astrophysics; Crystallography; Energy & Fuels; Imaging Science & Photographic Technology; Mechanics; Microscopy; Optics; Physics, Applied; Physics, Atomic, Molecular & Chemical; Physics, Condensed Matter; Physics, Fluids & Plasmas; Physics, Mathematical; Physics, Multidisciplinary; Physics, Nuclear; Physics, Particles & Fields; Spectroscopy; Thermodynamics |
| Psychology | Psychology; Psychology, Applied; Psychology, Biological; Psychology, Clinical; Psychology, Developmental; Psychology, Experimental; Psychology, Multidisciplinary; Psychology, Psychoanalysis; Psychology, Social |